%% file: dressed_faraday.tex
\documentclass[aps,prl,reprint,superscriptaddress,floatfix]{revtex4-1}

\input{preamble}

\begin{document}

\title{Continuously observing a dynamically decoupled spin-1 quantum gas}

\author{R.\,P.~Anderson}
\author{M.\,J.~Kewming}
\author{L.\,D.~Turner}
\affiliation{School of Physics \& Astronomy, Monash University, Victoria 3800, Australia.}

\date{\today}

\begin{abstract}
We continuously observe dynamical decoupling in a spin-1 quantum gas using a weak optical measurement of spin precession.
Continuous dynamical decoupling aims to dramatically modify the character and energy spectrum of spin states to render them insensitive to parasitic fluctuations. 
Continuous observation measures this new spectrum in a single-preparation of the quantum gas.
The measured time-series contains seven tones, which spectrogram analysis parses as splittings, coherences, and coupling strengths between the decoupled states in real-time.
With this we locate a regime where a transition between two states is decoupled from magnetic field instabilities up to fourth order, complementary to the parallel work at higher fields by Trypogeorgos \etal (\href{https://arxiv.org/abs/1706.07876}{arXiv:1706.07876}).
The decoupled microscale quantum gas offers magnetic sensitivity in a tunable band, persistent over many milliseconds: the length scales, frequencies, and durations relevant to many applications, including sensing biomagnetic phenomena such as neural spike trains.
\end{abstract}

\maketitle

From Hahn echoes to dynamical decoupling, pulse sequences have been used to protect spin superpositions from inhomogeneities and parasitic fluctuations, prolonging quantum coherence and circumventing deleterious energy shifts~\cite{biercuk_optimized_2009,lange_universal_2010,bluhm_dephasing_2011}.
A complementary strategy is continuous dynamical decoupling: replacing the pulse sequence with an uninterrupted coupling of bare spin states yielding dressed spin eigenstates with new quantization direction, spectrum, and coupling.
The new spectrum protects against low-frequency fluctuations, while the new couplings admit band-tunable sensing~\cite{fanchini_continuously_2007}.
Continuous dynamical decoupling has been applied to nitrogen-vacancy centers~\cite{hirose_continuous_2012,loretz_radio-frequency_2013,cai_robust_2012,*cai_long-lived_2012,golter_protecting_2014} and superconducting qubits, and creates protected qubit~\cite{aharon_general_2013} and decoherence-free~\cite{facchi_quantum_2002,*facchi_unification_2004} subspaces.
Marrying continuous dynamical decoupling with weak continuous measurement could give rise to new forms of quantum sensing exploiting synchronous detection and feedback~\cite{vijay_stabilizing_2012}.

Here we demonstrate how a weak continuous measurement of spin precession can probe the spectrum of a continuously decoupled spin-1 quantum gas in a single experimental preparation (`shot').
Time-resolved Fourier spectroscopy of this measurement record reveals not only all dressed-state splittings and their relative immunity to noise but also dressed state coherences and coupling strengths.
Estimating the eigenspectrum of the multi-level dressed system brings into view a higher-order decoupling than exists in dynamically-decoupled two-level systems.
In this regime, one transition is only quartically-sensitive to noise, surviving much larger amplitude fluctuations than conventional quadratic decoupling.
Further, a new transition arises between states which are otherwise uncoupled, completing a cyclic coupling of all dressed states. 
This low-frequency magnetic stability combined with continuous detection is immediately applicable to band-tunable magnetometry~\cite{hirose_continuous_2012,loretz_radio-frequency_2013,ockeloen_quantum_2013,*horsley_frequency-tunable_2016} and experiments preparing delicate spin-entangled many-body states~\cite{stamper-kurn_spinor_2013}; whereas the unconventional cyclic coupling could be applied to emulation of frustrated quantum spin chains~\cite{mikeska_one-dimensional_2004}.

Atomic Zeeman states $\ket{m_z=-1,0,1}$ in a magnetic field $B_z \vect{e}_z$ can be decoupled from fluctuations in $B_z$ by applying a perpendicular radiofrequency (rf) field $B_{\text{rf}} \vect{e}_x \cos \omega_\text{rf} t$, oscillating at $\omega_\text{rf}$, tuned near the Larmor frequency $\omega_L \equiv (E_{m_z=-1}-E_{m_z=+1})/2\hbar$.
At low magnetic fields, the degeneracy of the composite spin-$1/2$ systems~\cite{majorana_atomi_1932} renders the spin-1 behavior identical to continuous dynamical decoupling in spin-$1/2$ systems.
The spin is quantized along $\vect{\Omega} = \Omega \, \vect{e}_x + \Delta \, \vect{e}_z$ in a frame rotating with the radiofrequency $\omega_{\text{rf}}$; the eigenvalues of $\ham_{\text{rwa}} = \Delta \hat{F}_z + \Omega \hat{F}_x$ are $m_x \hbar \sqrt{\Omega^2 + \Delta^2}$, where $\Delta = \omega_{\text{rf}}-\omega_L$ is the detuning, $\Omega = \gamma B_{\text{rf}} / 2$ is the Rabi frequency, and $\ket{m_x=-1,0,1}$ is the corresponding eigenstate at resonance $\Delta=0$.
Radiofrequency dressing induces an avoided crossing in the spectrum; whereas the bare state energies are linearly sensitive to magnetic field variations $\delta B_z$ ($\omega_L \approx \gamma B_z$ where $\gamma$ is the gyromagnetic ratio), the dressed energies are only quadratically sensitive near resonance.

The spin character and symmetries are otherwise unchanged: transverse magnetic fields oscillating near the splitting frequencies drive transitions between eigenstates.
In the dressed system this means relatively low-frequency (`ac') fields oscillating near the Rabi frequency, such as $B_{y,\text{ac}} \vect{e}_y \cos \Omega t$ and $B_{z,\text{ac}} \vect{e}_z \cos \Omega t$, drive transitions $\ket{m_x=-1} \leftrightarrow \ket{m_x=0}$ and $\ket{m_x=0} \leftrightarrow \ket{m_x=+1}$.
This is the basis for ac magnetometry~\cite{hirose_continuous_2012} of relatively low-frequency fields; and for concatenated dynamical decoupling  which protects against fluctuations in $\Omega$~\cite{cai_robust_2012}.
Insensitivity to wider bandwidth and larger amplitude $\delta B_z$ can be achieved by increasing $\Omega$, opening a broader gap in the dressed spectrum, but doing so changes the detection band of ac magnetometry~\cite{loretz_radio-frequency_2013} or pulsed dynamical decoupling~\cite{boss_quantum_2017,*schmitt_submillihertz_2017}.
Henceforth we presume $\Omega$ is fixed by the application.

Any $\hat{F}_z^2$ interaction -- from nonlinear Zeeman~\cite{ramsey_molecular_1956}, microwave ac-Stark~\cite{gerbier_resonant_2006}, or tensor light~\cite{smith_continuous_2004} shifts -- raises the degeneracy of the $\ket{m_z=-1} \leftrightarrow \ket{m_z=0}$ and $\ket{m_z=0} \leftrightarrow \ket{m_z=+1}$ transitions. 
Now $\ham_{\text{rwa}} = \Delta \hat{F}_z + \Omega \hat{F}_x + q \hat{F}_z^2/\hbar$, where the quadratic Zeeman shift $q \equiv (E_{m_z=+1} + E_{m_z=-1} - 2 E_{m_z=0})_{\Omega=0}/2\hbar$.
\begin{figure}
    \includegraphics[width=\columnwidth]{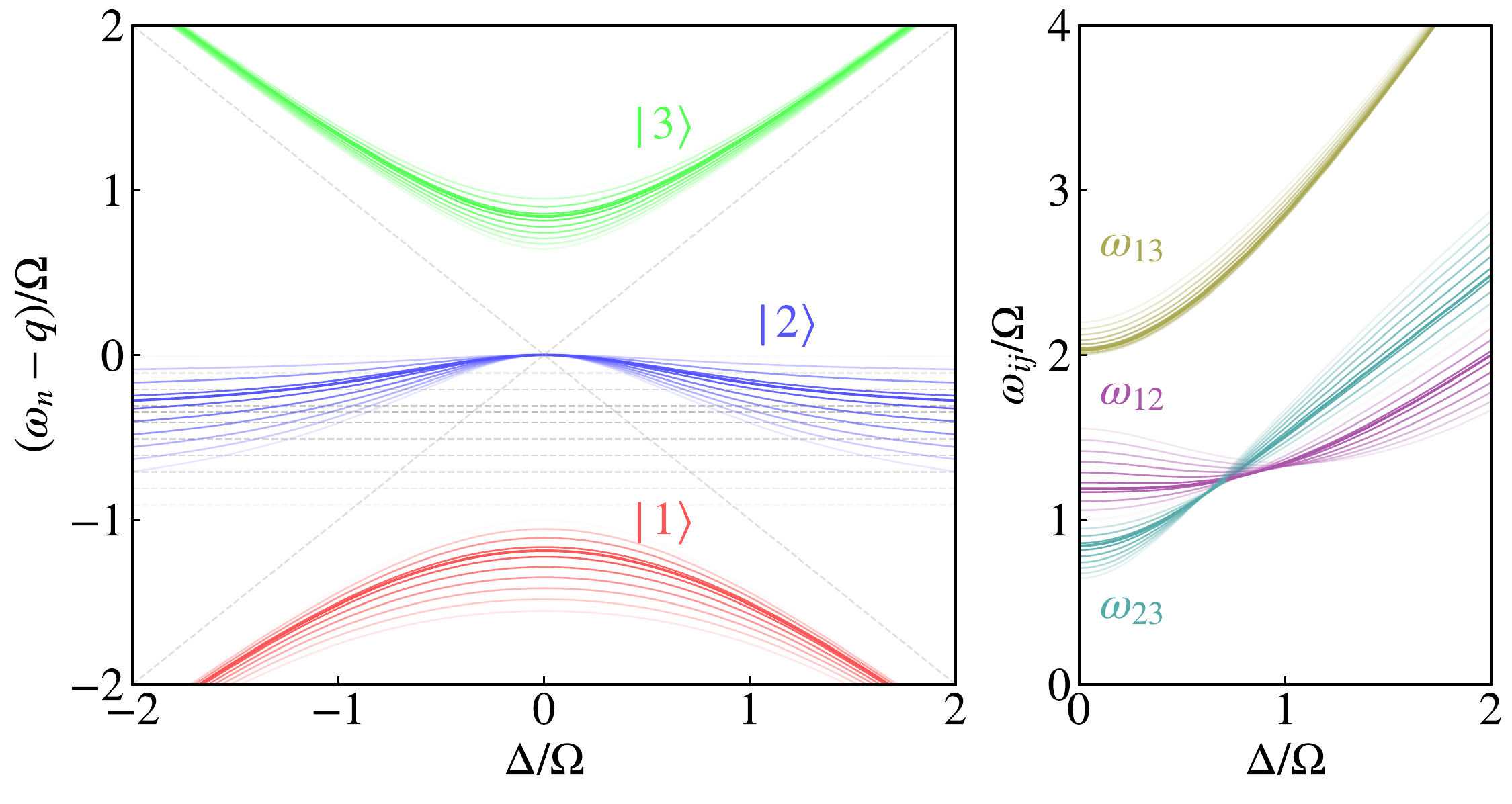}
    \caption{
    \label{fig:eigensystem_schematic}
        Energy spectrum and splittings of a radiofrequency coupled spin-1 for several $q_R = q/\Omega$ between $0$ and $1$.
        The transparency of each curve is proportional to the distance of $q_R$ from $\qRmagic$ in \refeq{eq:qRmagic}.
        (Left) Energies $\omega_n$ of dressed states $\ket{n}=\ket{1}$, (red) $\ket{2}$ (blue), and $\ket{3}$ (green) normalized to the Rabi frequency $\Omega$ as a function of detuning $\Delta$,
        Dashed lines indicate the energies of uncoupled states ($\Omega=0$) in a frame rotating at $\omega_{\text{rf}}$.
        (Right) Splittings $\omega_{ij} = \omega_j - \omega_i$ of dressed states $\ket{i}$ and $\ket{j}$ as a function of detuning.
        When $q_R=\qRmagic$ (bold curves), energies $\omega_1$ and $\omega_2$ share the same curvature, and their difference $\omega_{12}$ is minimally sensitive to detuning and thus magnetic field variations.
    }
\end{figure}
This yields dressed eigenstates $\{\ket{1}, \ket{2}, \ket{3}\}$ that are no longer $\text{SO}(3)$ rotations of the bare Zeeman states $\ket{m_z}$, and an eigenspectrum $\omega_i(\Delta) = E_{\ket{i}}/\hbar$ shown in \reffig{fig:eigensystem_schematic} (left).

Moreover, the couplings between these dressed states when $q \neq 0$ are markedly different:
$\bra{1} \hat{F}_{y,z} \ket{2}$ and $\bra{2} \hat{F}_{y,z} \ket{3}$ remain non-vanishing but $\bra{1} \hat{F}_x \ket{3}$ becomes non-zero.
The transitions are thus cyclic ($\ket{1} \leftrightarrow \ket{2} \leftrightarrow \ket{3} \leftrightarrow \ket{1}$) and non-degenerate, characterized by a dressed Larmor frequency $\omega_D\equiv(\omega_3-\omega_1)/2$, and dressed quadratic shift $q_D \equiv (\omega_3 + \omega_1 -2\omega_2)/2$, giving splittings $\omega_{23}=\omega_D-q_D$, $\omega_{12}=\omega_D+q_D$ and $\omega_{13}=2\omega_D$.
On resonance, $\omega_D=\sqrt{\Omega^2+q_D^2}$ and $q_D = -q/2$.

A figure-of-merit for decoupling is the curvature of the transition frequency at resonance.
In a dressed two-level system there is one convex and one concave eigenstate whose splitting is simply convex.
Figure~\ref{fig:eigensystem_schematic} shows that in the spin-1 system with quadratic shift, two states are convex.
This suggests that a regime may exist in which the curvature of their transition frequency vanishes~\cite{rabl_strong_2009,*xu_coherence-protected_2012}.
Indeed we find an analytic value of the normalized quadratic shift $q_R=q/\Omega$ where the curvatures of $\omega_1$ and $\omega_2$ are equal~\footnote{
  The curvature of the dressed-state energies is evaluated using perturbation theory. In particular, the dimensionless curvature of $\omega_{12}$ is $\partial^2(\omega_{12}/\Omega)/\partial(\Delta/\Omega)^2 = \Omega \, \partial^2\omega_{12}/\partial \Delta^2 = -(3 q_R \sqrt{4 + q_R^2} - q_R^2 - 2)/\sqrt{4 + q_R^2}$. For $q_R = 0$, we recover the spin-1/2 result, $\Omega\, \partial^2\omega_{12}/\partial \Delta^2 = 1$.},
\begin{align}
\label{eq:qRmagic}
    \qRmagic = \sqrt{(3\sqrt{2} - 4)/2} \approx 0.348 \, ,
\end{align}
resulting in the vanishing quadratic dependence of the transition frequency $\omega_{12}=\omega_2 - \omega_1$ on $\Delta$.
The leading-order sensitivity of these states to field variations $\delta B_z$ at $\qRmagic$ is quartic~\footnote{
    We take $\Delta = -\gamma \, \delta B_z$ for $|\Delta | \leq 2\Omega$ ($| \delta B_z | \leq B_{\text{rf}}/2$) and $| \partial q / \partial \Delta | \approx | \gamma^{-1} \partial q / \partial B_z | = |2 B_z q_Z / \gamma| \ll 1$, valid to $10^{-3}$ for the field strengths $B_z \lesssim \unit[5]{G}$ used here, resulting in vanishing third-order derivatives of $\omega_i$ with respect to detuning. 
    In general, the variation of $q$ with $\Delta$ (or $\delta B_z$) can be accounted for using the Breit-Rabi equation, leading to a residual linear and cubic variation of $\omega_{12}$ with $\delta B_z$~\cite{trypogeorgos_synthetic_2017}.},
giving the subspace comprised of ${\ket{1},\ket{2}}$ a higher-order decoupling than can be achieved with a two-level system; we term these the `hyper-decoupled states'.

We explore this high-order decoupling in the laboratory with a continuous measurement of the dressed spectrum of a spin-1 non-degenerate quantum gas.
Using a single realization of the quantum gas we make many successive weak measurements, revealing all three splittings $\omega_{ij}$ simultaneously.
Our spinor quantum gas apparatus~\cite{wood_magnetic_2015} and Faraday atom-light interface~\cite{jasperse_magic-wavelength_2017} are described elsewhere.
We prepare an ultracold gas ($\sim \unit[1]{\upmu K}$) of approximately $10^6$ \Rb atoms in a crossed-beam optical dipole trap.
A radiofrequency field of amplitude $\Omega/2\pi \leq \unit[100]{kHz}$ couples the three Zeeman states $\ket{m_z=-1,0,+1}$ of the lowest hyperfine ground state.
To perform a weak measurement of the evolving spin, we focus onto the atoms a linearly polarized far-off-resonant probe beam ($\lambda=\unit[781.15]{nm}$, red detuned $\unit[0.45]{THz}$) propagating along $x$.
The spin component parallel to the wavevector of the probe rotates its polarization via the paramagnetic Faraday effect; shot-noise limited polarimetry measures $\expect{\hat{F}_x}$ as an modulated tone near $\omega_L$.
Similar weak continuous Faraday measurements have tracked spin-mixing dynamics of a polar spinor condensate~\cite{liu_quantum_2009} and enabled quantum state tomography~\cite{smith_continuous_2004,smith_efficient_2006}.

To probe the dressed state spectrum and coherences, we prepare a superposition of dressed states by suddenly turning on the Rabi coupling $\Omega$, projecting the polarized collective spin $\ket{m_z=-1}$ onto $\ket{\psi(t=0)} = \sum_i c_i \ket{i}$.
The total magnetic field in the laboratory frame is $\vect{B}(t \geq 0) = -B_{\text{rf}} \cos (\omega_{\text{rf}} t) \vect{e}_x + B_z(t) \vect{e}_z$, where $B_z(t)$ varies slowly compared to $\Omega$.
The resulting Faraday signal is analyzed in the time-frequency domain using the short-time Fourier transform (STFT), revealing the rich frequency and amplitude modulation related to the dressed state energies, coherences, and coupling strengths.
Weak measurement allows this Fourier transform spectroscopy to be performed many times in one shot; recently Fourier transform spectroscopy has been applied to a quantum gas using projective measurements over many shots~\cite{valdes-curiel_fourier_2017}.

\begin{figure}
    \includegraphics[width=\columnwidth]{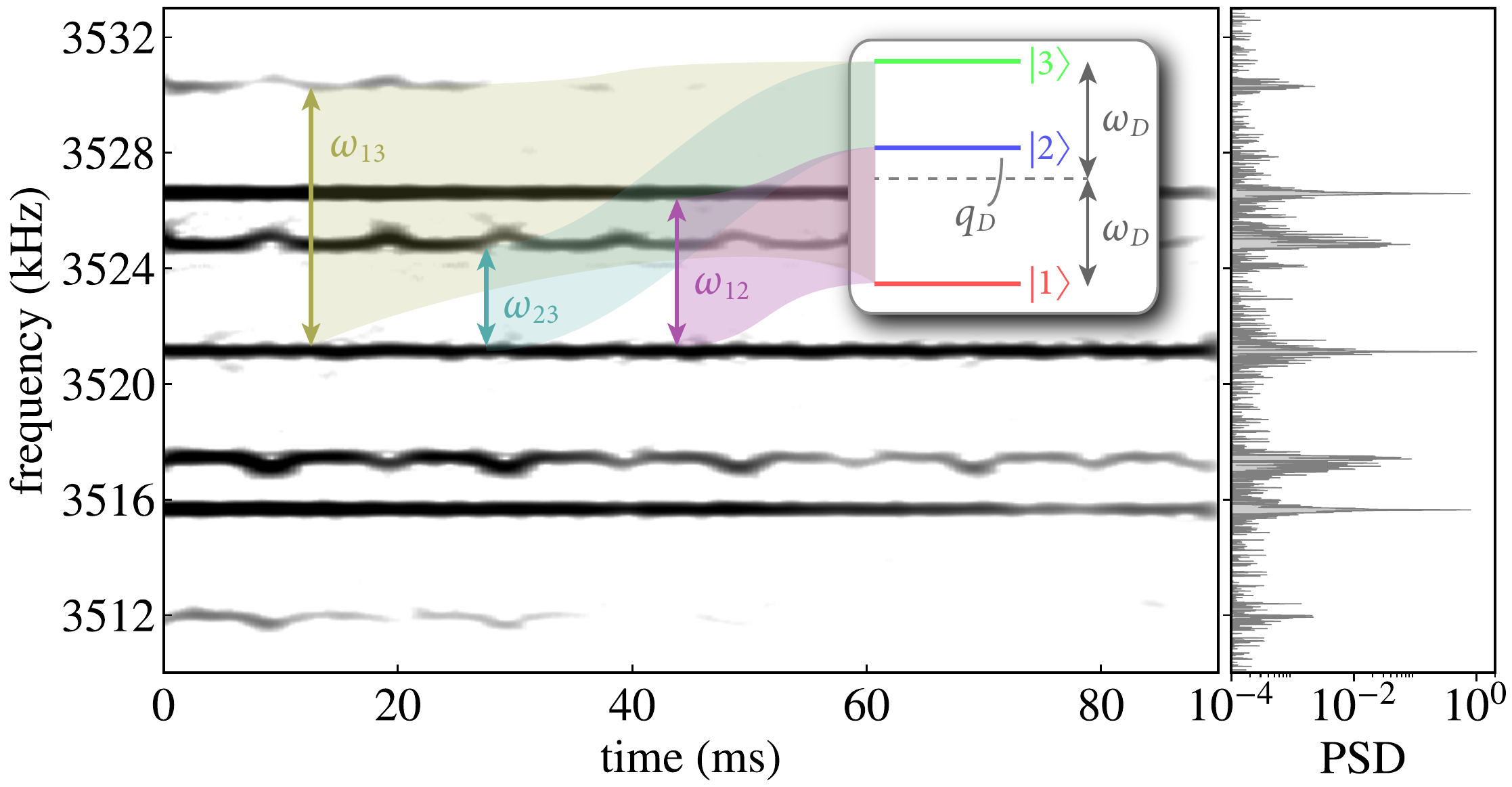}
    \caption{
    \label{fig:static_coupling}
    Continuous measurement of the dressed energy spectrum for $q_R = 0.402(3)$, $f_{\text{rf}}=\unit[3.521]{MHz}$ and $B_0=\unit[5.013]{G}$.
    (Left) Spectrogram and (right) power spectral density (PSD, normalized) of the $\unit[90]{ms}$ long signal. 
    (Inset) The dressed state energy diagram; the mean and difference of transition frequencies $\omega_{12}$ and $\omega_{23}$ is the dressed Larmor frequency $\omega_D$ and quadratic shift $q_D$, respectively.
    The sidebands about the carrier at $f_{\text{rf}}$ are associated with the $\omega_{13}$ (gold), $\omega_{23}$ (turquoise), and $\omega_{12}$ (lavender) transitions.
    Magnetic field fluctuations are manifest as asymmetric frequency modulation of the $\omega_{13}$ and $\omega_{23}$ sidebands, while the $\omega_{12}$ transition remains relatively unaffected.
    The corresponding peaks in the PSD have linewidths $\unit[102]{Hz}$, $\unit[97]{Hz}$, and $\unit[24]{Hz}$ (near transform-limited), respectively.
    The $\omega_{23}$ peaks are significantly skewed (mean magnitude of Pearson skew coefficient is $0.88$) while the $\omega_{12}$ peaks are unskewed (respectively $0.08$).
    }
\end{figure}

With no deliberate variation of the Rabi frequency or detuning, we observe the STFT amplitude (spectrogram) shown in \reffig{fig:static_coupling}.
Strong amplitude modulation of the Faraday signal is apparent as three pairs of sidebands, each equidistant from the carrier frequency $f_{\text{rf}} = \omega_{\text{rf}}/2\pi$. 
Each pair of sidebands corresponds to a dressed state transition $\ket{i} \leftrightarrow \ket{j}$; with sideband frequencies  $f_{\text{rf}} \pm f_{ij}$ where $f_{ij} = \omega_{ij}/2\pi$.
Thus the spectrogram is a calibration-free, real-time measurement of the dressed state spectrum.
Restricting attention to the upper sidebands, the two closest to the carrier are from adjacent state transitions $\omega_{12}$ and $\omega_{23}$ with similar amplitudes and at frequencies $\omega_D \pm q_D$ above the carrier.
The third, weaker sideband $2\omega_D$ above the carrier signifies the cyclic $\ket{1} \leftrightarrow \ket{3}$ transition, appearing when $q\neq 0$. 
No attempt is made to shield the apparatus from magnetic noise.
The power line causes a temporally varying $\delta B_z = \delta B_{\text{line}}(t)$ at the line frequency of $\unit[50]{Hz}$ and its odd harmonics, of $\unit[3.6]{mG}$ ($\unit[2.5]{kHz}$) peak-to-peak amplitude.
Each dressed transition is affected by the magnetic fluctuations differently: the sidebands corresponding to the $\omega_{13}$ and $\omega_{23}$ transitions exhibit asymmetric frequency modulation, whereas the optimally decoupled $\omega_{12}$ transition remains unperturbed within the frequency resolution of this spectrogram.
The normalized power spectral density (\reffig{fig:static_coupling}, right) of the entire time-series yields maximum frequency resolution at the expense of all temporal resolution.
The $\omega_{12}$ transition has a near transform-limited width, four times narrower than the $\omega_{23}$ and $\omega_{13}$ sidebands, which are also outwardly-skewed as expected transitions convex in $\Delta$. 
\begin{figure}
    \includegraphics[width=\columnwidth]{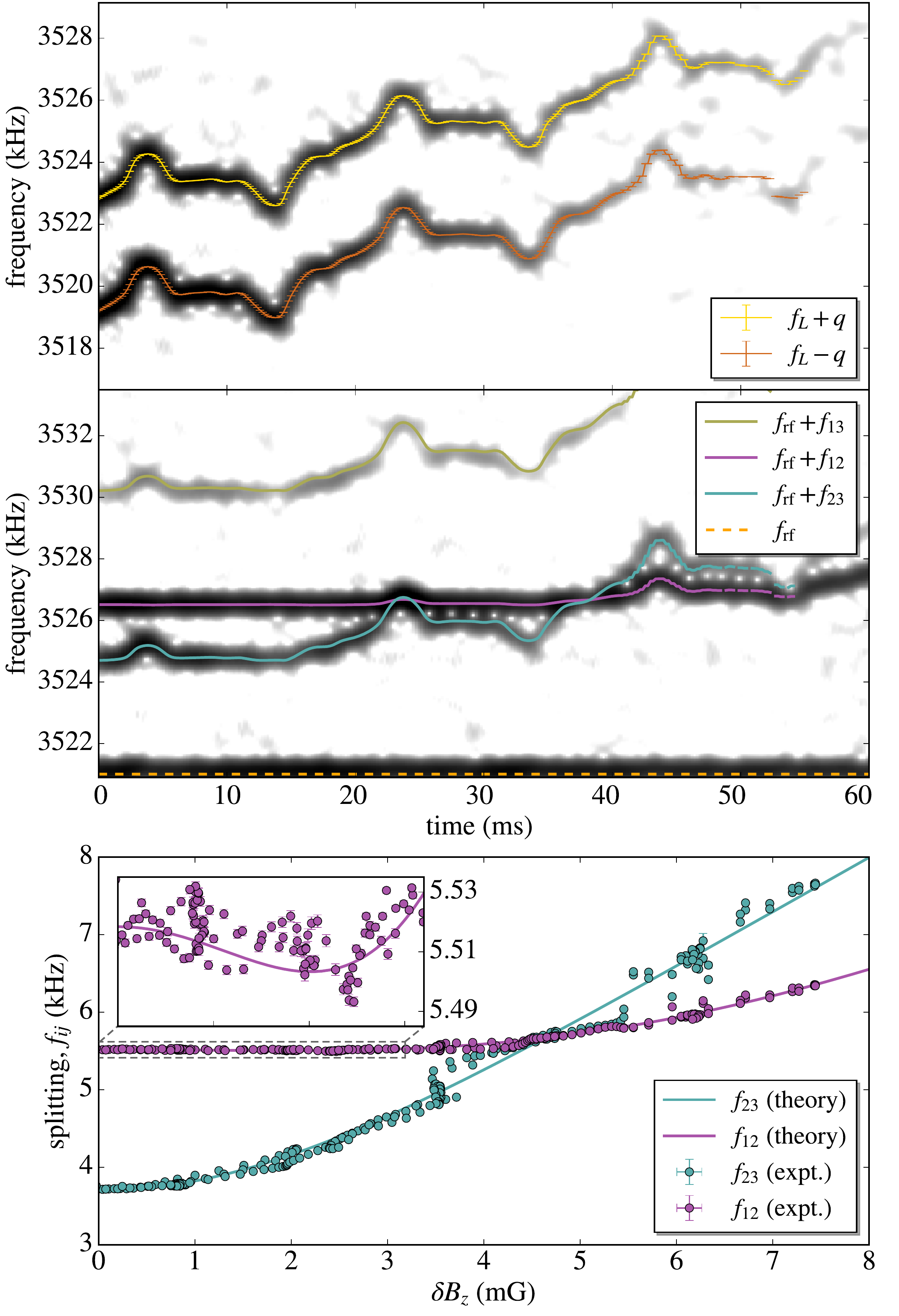}
    \caption{
    \label{fig:acquisition_pipeline}
        Real-time observation of continuous dynamical decoupling for $q_R = 0.402(3)$.
        (a) and (b) are spectrograms of a continuous weak measurement of $\expect{\hat{F}_x}$.
        (a) Magnetometry of the bare Zeeman states ($\Omega=0$) is used to calibrate $B_z(t) = B_0 + \delta B_z(t)$ during the measurement interval, in which the field [detuning] varies over a range $\sim B_{\text{rf}}$ [$2\Omega$].
        We numerically track the Zeeman splittings (gold/orange) to determine the instantaneous Larmor frequency $\omega_L(t)$ and quadratic shift $q(t)$.
       (b) The field is swept over the same range but the rf dressing is applied ($\Omega/2\pi = \unit[4.520(2)]{kHz}$).
       Three sidebands above (shown) and below the carrier at $f_{\text{rf}}=\unit[3.521]{MHz}$ (dashed, orange) reveal the dressed state splittings $f_{ij} = \omega_{ij}/2\pi$.
       (c) A parametric plot of $f_{12}(t)$ and $f_{23}(t)$ versus $\delta B_z(t)$ by combining analysis of (a) and (b).
       Solid curves in (b) and (c) are theoretical splittings from an eigenspectrum calculation, provided only $f_{\text{rf}}$, $B_z(t)$, and $\Omega$, i.e. no free parameters.
       Variation of the hyper-decoupled transition $f_{12}$ for $0 \leq \delta B_z \leq B_{\text{rf}}/4 = \unit[3.2]{mG}$ (c, inset).
    }
\end{figure}

The amplitude of each sideband is proportional to the corresponding dressed-state coherence $\rho_{ij} = c_i^* c_j $, and the non-vanishing dressed state coupling(s) $\bra{i} \hat{F}_{x,y,z} \ket{j}$.
\begin{table*}[t]
    \caption{Upper sidebands of the Faraday rotation signal $\propto \expect{\hat{F}_x}$ of an arbitrary dressed state superposition.
    Each sideband is identified with a dressed-state transition $\ket{i} \leftrightarrow \ket{j}$.
    Sideband frequencies are reported in both absolute terms and relative to the carrier at $\omega_{\text{rf}}$.
    Sideband amplitudes are for resonant coupling ($\Delta = 0$), and
    for the initial state $\ket{\psi(t=0)}=\ket{m_z=-1}$ these can be concisely expressed in terms of the dressed Larmor frequency $\omega_D$ and quadratic shift $q_D$.
    For each upper sideband, there is a lower sideband of the same amplitude, relative frequency, and opposite phase.
    \label{tab:sidebands}
    }
    \begin{ruledtabular}
    \begin{tabular}{ccccc}
    transition & frequency & $\omega_{ij} - \omega_{\text{rf}}$ & amplitude ($\Delta = 0$) & amplitude ($\Delta = 0$, $\ket{m_z=-1}$) \\ \hline
     (carrier) & $\omega_{\text{rf}}$ & 0 & $(\bra{3} \hat{F}_x \ket{3} - \bra{1} \hat{F}_x \ket{1}) (\rho_{33} - \rho_{11})$  & $\hbar q_D \Omega/2 \omega_D^2$ \\
     $\ket{1} \leftrightarrow \ket{2}$ & $\omega_{\text{rf}} + \omega_{12}$ & $\omega_D+q_D$ & $-2i \bra{1} \hat{F}_y \ket{2} \text{Re}\,\rho_{12} = -2 \bra{2} \hat{F}_z \ket{3} \text{Re}\,\rho_{12}$ & $\hbar \Omega/4 \omega_D$ \\
     $\ket{2} \leftrightarrow \ket{3}$ & $\omega_{\text{rf}} + \omega_{23}$ & $\omega_D-q_D$ & $2i \bra{2} \hat{F}_y \ket{3} \text{Re}\,\rho_{23} = 2 \bra{1} \hat{F}_z \ket{2} \text{Re}\,\rho_{23}$ & $\hbar \Omega/4 \omega_D$ \\
     $\ket{1} \leftrightarrow \ket{3}$ & $\omega_{\text{rf}} + \omega_{13}$ & $2\omega_D$ & $2 \bra{1} \hat{F}_x \ket{3} \text{Re}\,\rho_{13}$ & $\hbar q_D \Omega/4 \omega_D^2$
    \end{tabular}
    \end{ruledtabular}
\end{table*}
Analytic expressions for the sideband amplitudes near resonance ($\abs{\Delta} \ll \Omega$) are summarized in Table~\ref{tab:sidebands}.
If the projection onto the dressed basis (and hence $\rho_{ij}$) is known, our measurement constitutes a single-shot estimation of the coupling strengths.  
Alternatively, if the couplings are separately characterized~\cite{trypogeorgos_synthetic_2017}, this amounts to continuous measurement of the dressed density matrix, effecting quantum state estimation of the dressed system.

Different platforms use different metrics for the fidelity of dynamical decoupling, and in addition to linewidth narrowing include prolonged coherence.
We observe a three-fold increase in the lifetime of the spectral components corresponding to the $\omega_{12}$ and $\omega_{23}$ transitions as compared with the undressed system (\reffig{fig:acquisition_pipeline}a, $1/e$ decay time \unit[23.8(2)]{ms}).
Dressed-state coherences are expected to last longer, but were limited here by the $\sim\unit[100]{ms}$ probe-induced photon scattering time.
A less perturbative probe~\cite{jasperse_magic-wavelength_2017} should reveal even longer dressed coherence times at the expense of signal-to-noise ratio. 

To better expose the enhanced insensitivity of the hyper-decoupled states in the vicinity of $\qRmagic$, we sweep the the magnetic field over a wider range than is furnished by the power line noise.
The longitudinal field $B_z(t) = B_0 + \alpha t + B_{\text{line}}(t)$, where $\alpha = \unit[128]{mG/s}$ is the linear sweep rate; the resulting detuning sweeps across $2\Omega$ (\textit{cf.} the domain of~\reffig{fig:eigensystem_schematic}) during the single-shot measurement.
We interleave each rf-dressed shot with a magnetometry shot calibrating $B_z(t)$: an rf $\pi/2$-pulse initiates Larmor precession of the undressed collective spin, and the Faraday signal is composed of two tones at $\omega_\pm = \omega_L \pm q$, the Zeeman splittings (\reffig{fig:acquisition_pipeline}, top).
For $q \, \tau_f \geq 2\pi$, where $\tau_f$ is the length of the spectrogram window, $\omega_\pm$ are resolved yielding the instantaneous $\omega_L(t)$ and $q(t)$.
We then use $\omega_L(t)$ to find $\delta B_z(t)$ (and $\Delta(t)$) by inverting the Breit-Rabi equation~\cite{ramsey_molecular_1956,Note3}.

We measure the dressed spectrum for resonant magnetic fields $B_0$ ranging from $3.549$ to $\unit[5.568]{G}$ (applied rf frequencies $f_{\text{rf}}$ from $2.493$ to $\unit[3.911]{MHz}$), with a mean Rabi frequency of $\Omega/2\pi = \unit[4.505(3)]{kHz}$ ($B_{\text{rf}} = \unit[12.83(1)]{mG}$).
At each field $B_0$ we ensure the Rabi frequency is fixed by measuring the voltage drop across the coil at $f_{\text{rf}}$ with an rf lock-in amplifier.
The Rabi frequency is ultimately measured using the atoms by analyzing the dressed energy spectrum near resonance ($\abs{\Delta}/2\pi \leq \unit[100]{Hz}$) where $\Omega = \sqrt{\omega_{12} \omega_{23}}$.
The measured Rabi frequencies have a standard deviation $\sigma(\Omega)/2\pi = \unit[9.4]{Hz}$, validating the above method.

Figure~\ref{fig:acquisition_pipeline} shows the dressed spectrum measured as $\delta B_z$ varies across a range $\sim B_{\text{rf}}$ during a single-shot.
The instantaneous dressed state splittings for all three transitions are predicted with no free parameters, and plotted atop the spectrogram data, showing excellent agreement with the measured sidebands.
Line noise renders $\delta B_z(t)$ neither linear nor monotonic.
By tracking the instantaneous peaks in the calibration and dressed spectrograms we plot $(\delta B_z(t), f_{ij}(t))$ parametrically, eliminating the line noise systematic.
The sensitivity of the $\ket{1} \leftrightarrow \ket{2}$ and $\ket{2} \leftrightarrow \ket{3}$ transitions to magnetic field variations is shown in \reffig{fig:acquisition_pipeline}c.
The hyper-decoupled transition is least sensitive; $f_{12}$ varies by $\unit[39]{Hz}$ (expt.), $\unit[26]{Hz}$ (theory) for $0 \leq \delta B_z \leq B_{\text{rf}}/4 = \unit[3.2]{mG}$ (\reffig{fig:acquisition_pipeline}c, inset).
Normalizing the variation to the Rabi frequency makes possible a comparison of the decoupling across platforms and ac magnetometry bandwidths. 
The normalized variation $\omega_{12}/\Omega = 8.6\times10^{-3}$ (expt.) and $5.8\times10^{-3}$ (theory) across a detuning range of half a Rabi frequency.
By comparison, the normalized variation for conventional decoupling ($q_R=0$) is $(\sqrt{5}-2)/2 \approx 0.118$; 14 (expt.) and 20 (theory) times higher than the variation.
Alternatively, the normalized variation of the $\ket{m_z=\pm1} \leftrightarrow \ket{m_z = 0}$ Zeeman transitions in the low-field limit is $0.5$; 58 (expt.) and 86 (theory) times higher than the variation in the hyper-decoupled transition frequency.

\begin{figure}
    \includegraphics[width=\columnwidth]{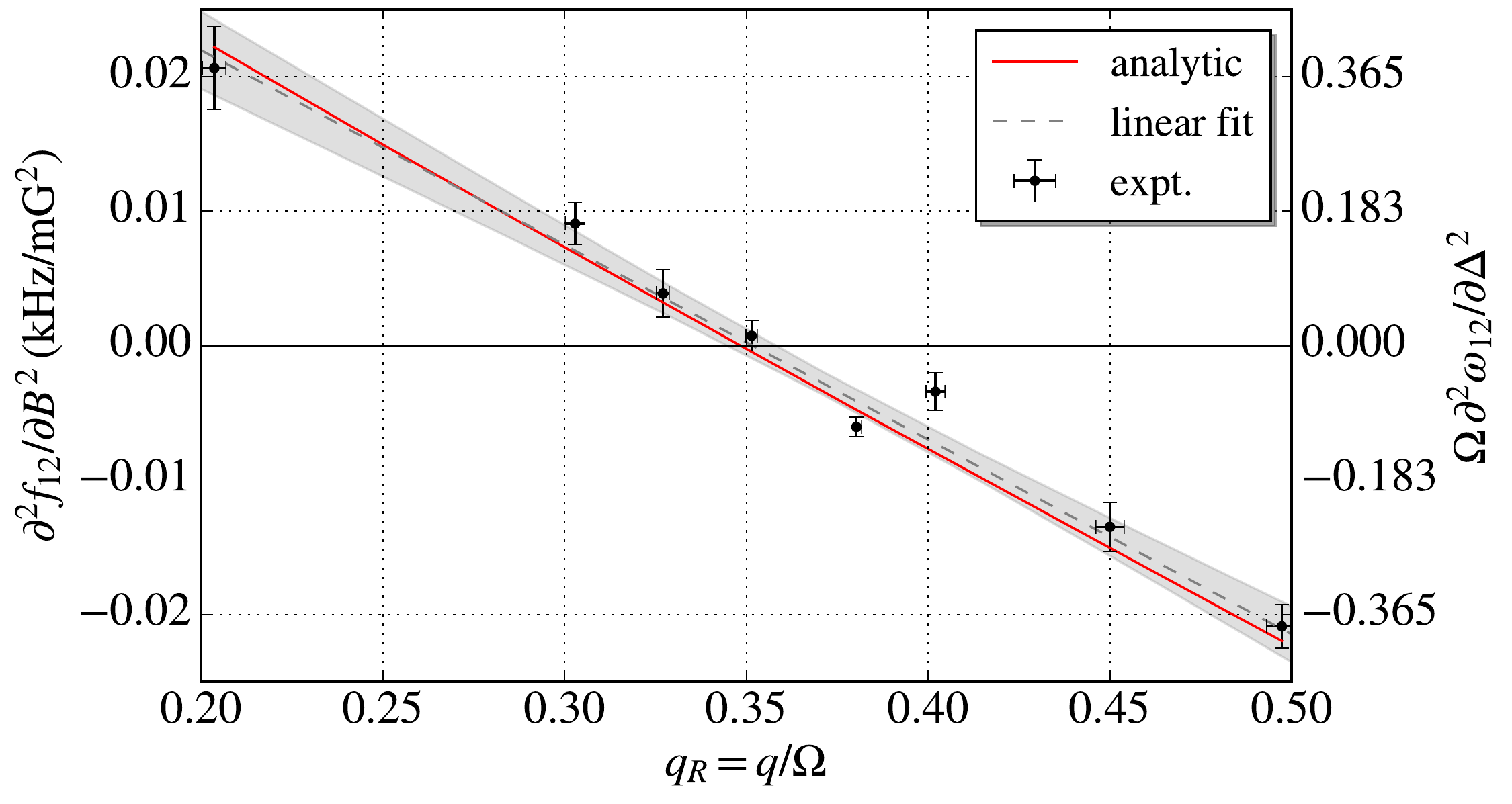}
    \caption{
    \label{fig:curvature_vs_qR}
        Curvature of the hyper-decoupled $\ket{1} \leftrightarrow \ket{2}$ transition for relative quadratic shifts $q_R$ from $0.2$ to $0.5$.
        The measured curvature (black points) is determined from polynomial fitting to $(\delta B_z, f_{12})$ data, e.g. \reffig{fig:acquisition_pipeline}(c).
        Vertical and horizontal error bars correspond to the standard error of the regression and uncertainty in $q_R$ (via $u(q)$ and $u(\Omega)$ at each field $B_0$), respectively.
        A linear fit (black, dashed) with 1$\sigma$ confidence band (gray, shaded) are shown, whose intercept can be used to impute $\qRmagic\text{ (expt.)} = 0.350(6)$.
        The analytic expression for the curvature (red)~\cite{Note1} is consistent with the data-driven analysis of the curvature, \textit{cf.} $\qRmagic\text{ (theory)} = 0.348$.
        The left [right] vertical axis shows the curvature $\partial^2 f_{12}/\partial B_z^2$ [$\Omega\, \partial^2\omega_{12}/\partial \Delta^2$] in absolute units of $\unit{kHz/G^2}$ [dimensionless units].
        The normalized curvature is unity when $q_R=0$.
    }
\end{figure}

To optimally suppress the sensitivity of the hyper-decoupled states to small field variations, we experimentally determine the curvature of $\omega_{12}$  for $q_R$ between 0.2 and 0.5, independent of the predicted spectrum of $\ham_{\text{rwa}}$.
For each $q_R$, we fit a polynomial to $(\delta B_z, f_{12})$ data to extract $\partial^2 f_{12}/\partial B_z^2$ (\reffig{fig:curvature_vs_qR}).
We perform linear regression of the measured curvature versus $q_R$ to infer $\qRmagic\text{ (expt.)} = 0.350(6)$.
The predictive power of the measured dressed spectrum and this model-independent analysis is affirmed by the agreement with the theoretical curvature (red curve, \reffig{fig:curvature_vs_qR}) and $\qRmagic$ in~\refeq{eq:qRmagic}.
The lowest curvature we measure is $\left(\partial^2 f_{12}/\partial B_z^2\right)_{\text{min}} = \unit[0.7]{Hz/mG^2}$ at $q_R = 0.351(2)$.
In dimensionless units with the splitting and detuning normalized to the Rabi frequency $\left(\Omega\, \partial^2\omega_{12}/\partial \Delta^2\right)_{\text{min}} = 0.013$, $75$ times lower than the curvature of this transition for quadratic decoupling ($q_R = 0$).

This intra-shot revelation of the time and frequency domain renders the measurement of these spectra orders of magnitude more efficient.
For example, the single spectrum shown in \reffig{fig:acquisition_pipeline} would take $\sim (10$ shots per $\delta B_z$ per $\omega_{ij} ) \times (20$ distinct $\delta B_z$) $\times (3 $ transitions $\omega_{ij}) = 600$ shots, or $\sim \unit[1.2\times10^4]{s} = 200$ minutes of data acquisition.
We acquired this spectrum in a single shot, i.e. $\unit[20]{s}$.
The data used to generate \reffig{fig:curvature_vs_qR} was acquired in only $5$ minutes.

In summary, we have demonstrated continuous measurement of continuous dynamical decoupling in a spin-1 quantum gas.
Continuous weak measurement via the Faraday effect yields information about the rf-dressed superposition, the dressed-state couplings and energies, simultaneously, making possible full characterization over detunings in a single experimental shot.
We posit that when viewed as an ac magnetometer, this information will not only measure fields oscillating at the dynamically-tunable Rabi frequency, but self-certifies both band center and residual detuning error, in real time, while remaining fourth-order decoupled.
More broadly, the cyclic coupling we observe may emulate quantum spin ladders with frustrated interactions~\cite{mikeska_one-dimensional_2004}.
Our measurement is readily extended into the back-action regime, where
measurement of bare precessing spins induces simultaneous two-axis squeezing~\cite{colangelo_simultaneous_2017}.
Strong measurement of dressed precessing spins adds the third axis and may expose non-Gaussian quantum noise geometries.

We thank A. A. Wood, I. B. Spielman, N. Lundblad, and F. A. Pollock for enlightening discussions.
This work was supported by ARC LP130100857 and a Monash University IDR seed grant. 
This work complements parallel measurements of the same rf-dressed system by Trypogeorgos \etal~\cite{trypogeorgos_synthetic_2017} using projective readout of dressed state populations at higher magnetic fields.

\bibliography{dressed_faraday}

\end{document}

%% file: preamble.tex
\usepackage{graphicx}
\usepackage[usenames,dvipsnames]{color}
\usepackage{amsmath,amssymb}
\usepackage{bm}
\usepackage{upgreek}
\usepackage{xspace}
\usepackage{units}
\usepackage{hhline}
\usepackage[vskip=0pt]{quoting}
\usepackage[colorlinks,urlcolor=blue,citecolor=blue,linkcolor=blue]{hyperref}
\graphicspath{{figures/}}

\newcommand{\etal}{et~al.\xspace}
\newcommand{\Rb}{$^{87}$Rb\xspace}

\newcommand{\qRmagic}{q_{R,\text{magic}}\xspace}

\newcommand{\reffig}[1]{\mbox{Fig.~\ref{#1}}}
\newcommand{\refeq}[1]{\mbox{Eq.~(\ref{#1})}}

\newcommand{\expect}[1]{\langle #1 \rangle}
\newcommand{\abs}[1]{\vert #1 \vert \xspace}
\newcommand{\bra}[1]{\langle #1 \vert \xspace}
\newcommand{\ket}[1]{\vert #1 \rangle \xspace}

\newcommand{\vect}[1]{\mathbf{#1}\xspace}

\newcommand{\ham}{\mathcal{H}}

\catcode`_=12
\begingroup\lccode`~=`_\lowercase{\endgroup\let~\sb}